# Motile *Geobacter* dechlorinators migrate into a model source zone of trichloroethene dense non-aqueous phase liquid: experimental evaluation and modeling

Jo Philips [a,1,*] (jo.philips@ugent.be)

Alexey Miroshnikov [b] (alexeym@cns.umass.edu)

Pieter Jan Haest [a,2] (pieterjan.haest@agt.be)

Dirk Springael [a] (dirk.springael@ees.kuleuven.be)

& Erik Smolders [a] (erik.smolders@ees.kuleuven.be)

[a] Department of Earth and Environmental Sciences,

Division of Soil and Water Management,

KU Leuven (University of Leuven),

Kasteelpark Arenberg 20, B-3001 Heverlee BELGIUM

Tel: +3216321609, Fax: +3216321997

[b] Department of Mathematics and Statistics,

University of Massachusetts,

Lederle Graduate Research Tower, Amherst MA 01003-9305, USA

* Corresponding author

[1] Current address: Laboratory for Microbial Ecology and Technology,

Ghent University, Coupure Links 653A, B-9000 Gent, BELGIUM

[2] Current address: A.G.T. n.v., Kontichsesteenweg 38, B-2630 Aartselaar, BELGIUM

**ABSTRACT**

Microbial migration towards a trichloroethene (TCE) dense non-aqueous phase liquid (DNAPL) could facilitate the bioaugmentation of TCE DNAPL source zones. This study characterized the motility of the *Geobacter* dechlorinators in a TCE to *cis*-dichloroethene dechlorinating KB-1$^{TM}$ subculture. No chemotaxis towards or away from TCE was found using an agarose in-plug bridge method. A second experiment placed an inoculated aqueous layer on top of a sterile sand layer and showed that *Geobacter* migrated several centimeters in the sand layer in just 7 days. A random motility coefficient for *Geobacter* in water of $0.24 \pm 0.02$ cm$^2$ day$^{-1}$ was fitted. A third experiment used a diffusion-cell setup with a 5.5 cm central sand layer separating a DNAPL from an aqueous top layer as a model source zone to examine the effect of random motility on TCE DNAPL dissolution. With top layer inoculation, *Geobacter* quickly colonized the sand layer, thereby enhancing the initial TCE DNAPL dissolution flux. After 19 days, the DNAPL dissolution enhancement was only 24% lower than with an homogenous inoculation of the sand layer. A diffusion-motility model was developed to describe dechlorination and migration in the diffusion-cells. This model suggested that the fast colonization of the sand layer by *Geobacter* was due to the combination of random motility and growth on TCE.

**HIGHLIGHTS**

- *Geobacter* dechlorinators are motile, but not chemotactic to TCE
- *Geobacter* dechlorinators quickly colonized model TCE DNAPL source zone and initiated bio-enhanced DNAPL dissolution
- Diffusion-motility model indicates that fast colonization is due to combination of random motility and growth on TCE

# 1 INTRODUCTION

Trichloroethene (TCE) has become a widespread groundwater pollutant, as a result of its extensive use as a solvent in industry. Remediation of aquifers contaminated with this chemical is challenging, since TCE forms dense non-aqueous phase liquids (DNAPL) in the subsoil. Bioremediation can be used to clean up such contaminations, as several anaerobic bacteria grow with the reductive dechlorination of TCE to *cis*-dichloroethene (*cis*-DCE) (Bradley, 2003). In the vicinity of a chlorinated ethene DNAPL, this dechlorination reaction can enhance the DNAPL dissolution flux and, as such, result in a reduced remediation time (Yang and McCarty, 2000). Dechlorination in a contaminated aquifer can be slow or lacking and, in such cases, bioaugmentation, i.e. the injection of a dechlorinating culture, can initiate bio-enhanced DNAPL dissolution (Adamson et al., 2003; Sleep et al., 2006). However, the low-permeability zones formed by DNAPLs are hard to reach if bacterial transport only relies on the groundwater flow (Singh and Olson, 2008). Microbial migration towards a DNAPL could largely accelerate the bioaugmentation of DNAPL source zones (Ford and Harvey, 2007; Singh and Olson, 2008). Microbial migration can be due to random motility, i.e. swimming of the bacterial cells in random directions. In addition, some bacteria are positive or negative chemotactic, i.e. they direct their swimming respectively towards or away from a chemical gradient. Both random motility and positive chemotaxis could increase the migration towards a DNAPL, whereas negative chemotaxis could prevent the exposure to the toxic TCE concentrations adjacent to a TCE DNAPL (Ford and Harvey, 2007; Singh and Olson, 2008).

Motility and chemotaxis have been demonstrated for aerobic bacteria degrading various environmental pollutants (Pandey and Jain, 2002; Singh and Olson, 2008). The aerobe *Pseudomonas putida* G7, for instance, is chemotactic towards naphthalene and was found to enhance the dissolution of this compound (Marx and Aitken, 2000; Law and Aitken, 2003). Chemotaxis towards TCE was observed for the aerobe *Pseudomonas putida* F1, which cometabolically converts TCE (Parales et al., 2000), whereas another study reported that this species migrated away from a dissolving TCE DNAPL (Singh and Olson, 2010). Among the anaerobic contaminant degraders, motility was observed for several perchloroethene (PCE) and TCE dechlorinating bacteria, including *Geobacter lovleyi* SZ (Holliger et al., 1998; Sung et al., 2006). Bacteria highly related to the latter species are responsible for the TCE to *cis*-DCE dechlorination by the commercially available KB-1$^{TM}$ culture often used for bioaugmentation (Duhamel and Edwards, 2007; Philips et al., 2012). So far, however, it remains unknown whether dechlorinating bacteria are able to migrate into a TCE DNAPL source zone and can initiate bio-enhanced DNAPL dissolution. Therefore, the aim of this study was to investigate the motility and potential chemotaxis of *Geobacter* dechlorinators and to examine the effect of their migrational properties on the dissolution of a TCE DNAPL.

All experiments used a TCE to *cis*-DCE dechlorinating KB-1$^{TM}$ subculture, containing dechlorinators highly similar to *Geobacter lovleyi* SZ (Philips et al., 2012). Firstly, a chemotaxis assay was conducted, but no evidence for chemotaxis by the culture towards or away from TCE was found. A second experiment quantified the random motility of the *Geobacter* dechlorinators by sampling an initially sterile sand layer, on top of which an inoculated aqueous layer was placed. A third experiment investigated the effect of the motility of the *Geobacter* dechlorinators on the dissolution of a TCE DNAPL. Hereto, model source zones were created using our previously described diffusion-cell setup (Philips et al., 2011). Finally, a diffusion-motility model was used to clarify how the motility of *Geobacter* contributed to the dechlorination observed in the diffusion-cell experiment.

## 2 MATERIAL AND METHODS

### 2.1 Medium and culture

Anaerobic defined medium was prepared as described by Haest et al. (2011), except that the concentration of yeast extract was lowered to 10 mg·L$^{-1}$. Briefly, this medium uses 30 mM 3-(N-morpholino)propanesulfonic acid (MOPS, pKa =7.2) and 15 mM hydroxide to buffer the pH, while 1 mM cysteine was used as reducing agent. A KB-1$^{TM}$ subculture dechlorinating TCE to *cis*-DCE with formate as electron donor was used as inoculum in all experiments. This KB-1$^{TM}$ subculture was grown on 1.6 mM TCE to stimulate acclimatization to high TCE concentrations, while only TCE to cis-DCE conversion was stimulated to obtain high TCE dechlorination rates. Formate was chosen as electron donor because of its good pH buffering capacity (Philips et al., 2013b) and was added in an amount equivalent for the conversion of TCE to *cis*-DCE. The composition of this KB-1$^{TM}$ subculture was described previously (Philips et al., 2012). In general, this culture consists of dechlorinating species highly similar to *Geobacter* *lovleyi* SZ and fermenting species related to *Clostridium,* while *Dehalococcoides* dechlorinators were outcompeted. Dechlorination by this culture occurs at TCE concentrations below 2.6 mM (Philips et al., 2013a).

### 2.2 Chemotaxis assay

Chemotaxis towards or away from TCE was assessed using the agarose-in-plug bridge method (Yu and Alam, 1997). Pure TCE liquid was stained with 0.5 g·L$^{-1}$ Oil-Red-O and several droplets were mixed with a warm solution of 2% low-melting-temperature agarose. 10 µl of this mixture was placed on a microscope slide and a cover slip supported by two other cover slips was immediately placed over the drop. Control plugs were made without the addition of TCE liquid. The KB-1$^{TM}$ subculture was flooded around the plug inside a glovebox with $N_2/H_2$ (95/5) atmosphere. The slides were incubated up to 30 min. The plugs were visualized using dark field microscopy.

### 2.3 Motility experiment

The motility of the KB-1$^{TM}$ subculture in a porous medium was studied using two 12 cm long glass cylinders with a 3.3 cm internal diameter and sampling ports spaced about 5 mm lengthwise (Philips et al., 2011). These cylinders contained a 9 cm sand layer with on top a 3 cm aqueous layer to which the inoculum was added. The sand layer was created by mixing sterile anaerobic medium with sterile sand in a glovebox with $N_2/H_2$ (95/5) atmosphere. The sand was of the same type as described before (Philips et al., 2011) and was sterilized beforehand by autoclaving a slurry of sand and deionized water and drying it at 60°C. A dense and homogeneous packing of the sand layer was obtained by stirring the sand and tapping the glass with a rubber stick. The top layer (25 mL) consisted of anaerobic medium inoculated with 60 volume % of the KB-1$^{TM}$ subculture, which corresponded with respectively $8 \cdot 10^7$ and $2 \cdot 10^8$ *Geobacter* and bacterial 16S rRNA gene copies per mL, as determined with quantitative PCR (qPCR) (described below). No TCE or formate was added. Incubation was performed at 20 °C. One cylinder was sampled four days after the start of the experiment, whereas the other cylinder was sampled after seven days. Pore water samples were taken for DNA analysis.

### 2.4 Diffusion-cell experiment

The effect of the motility of *Geobacter* dechlorinators on TCE DNAPL dissolution was studied in model source zones using a three-layer diffusion-cell setup (Philips et al., 2011). This setup uses the same glass cylinders as described above, in which 3 layers are created, i.e. a central 5.5 cm sand layer, a lower 3.5 cm DNAPL layer and an upper 3 cm aqueous layer. The DNAPL layer is an aqueous solution containing pure TCE droplets and is continuously stirred to maintain the saturated TCE concentration. The aqueous top layer is frequently replaced with fresh medium to provide electron donor and to remove the chloroethenes.

All diffusion-cells were filled in a glovebox with $N_2/H_2$ (95/5) atmosphere using the same type of sand as described previously. To allow comparison with previous experiments, a first set of diffusion-cells was inoculated in the sand layer. In these diffusion-cells, sterile sand was homogenously mixed with anaerobic medium inoculated with 0.6, 6 or 60 volume % of the KB-1$^{TM}$ subculture. These inoculation densities corresponded with $4 \cdot 10^5$, $4 \cdot 10^6$ or $4 \cdot 10^7$ *Geobacter* 16S rRNA gene copies per mL and $1 \cdot 10^6$, $1 \cdot 10^7$ or $1 \cdot 10^8$ bacterial 16S rRNA gene copies per mL, as determined with qPCR (described below). Treatments were performed in triplicate (0.6%) or in duplicate (6 and 60%). Top layers were refreshed twice a week with fresh medium containing 4 mM formate.

A second set of diffusion-cells investigated the effect of the motility of the *Geobacter* dechlorinators on TCE DNAPL dissolution by inoculating the top layer. Before inoculation, these diffusion-cells were operated under abiotic conditions for three weeks. During this period, top layers were replaced twice a week with fresh medium. Measurement of the pore water TCE concentrations showed that at the end of this period the TCE concentration linearly declined through the sand layer from the saturation TCE concentration (8.4 mM) at the DNAPL layer to an about zero concentration at the top layer (results not shown). We previously obtained a similar TCE concentration profile and explained that this linear profile is the expected steady-state in abiotic conditions (Philips et al., 2011). After abiotic steady-state TCE concentrations were established, the top layer of these diffusion-cells was inoculated with 0.6, 6 or 60 volume % of the KB-1$^{TM}$ subculture. These inoculation densities correspond with the cell densities given above. Treatments were performed in duplicate (0.6 and 6%) or triplicate (60%). The top layer was not refreshed during the first week after inoculation to allow microbial migration towards the sand layer. Afterwards, top layers were again refreshed twice a week with fresh medium. The top layers were amended with 4 mM formate as electron donor, once the biotic conditions commenced.

Incubation of all diffusion-cells was performed at 20°C. Pore water samples were taken 19 days after inoculation. Samples of duplicate diffusion-cells were taken for analysis of the chlorinated ethene concentrations, whereas the remaining triplicate diffusion-cells were sampled for DNA analysis.

## 2.5 Analytical methods

Pore water concentrations of TCE and *cis*-DCE were analyzed using GC-FID (Haest et al., 2010a). Stock solutions of these compounds in methanol were used as external standards. The DNeasy Blood and Tissue kit (Qiagen) was used to extract DNA from pore water samples (Philips et al., 2012) and *Geobacter* and bacterial 16S rRNA gene copy numbers were quantified with real-time quantitative PCR (qPCR) (Haest et al., 2011). An aqueous control sample was included in each combined DNA extraction and qPCR run, for reasons described previously (Philips et al., 2012). The obtained 16S rRNA gene copy numbers correspond with cell densities (Philips et al., 2012). The detection limit for the qPCR analysis was $1 \cdot 10^4$ 16S rRNA gene copies per mL. Restriction Fragment Length Polymorphism (RFLP) was performed on amplified bacterial 16S rRNA gene fragments as previously described (Philips et al., 2012).

## 2.6 Calculation of the random motility coefficient

The experimental results of the motility experiment were used to calculate a random motility coefficient for *Geobacter*. Random motility is mathematically described similarly to solute diffusion (Ford and Harvey, 2007). As such, random motility coefficients are analogues of diffusion coefficients. Therefore, Equation 1 was adapted from the analytical solution for solute diffusion in a semi-infinite region (Datta, 2002), assuming a zero initial *Geobacter* cell density in the sand layer and a constant *Geobacter* cell density at the top:

$$c_{geo} = c_{geo,top} \cdot \left(1 - erf\left(\frac{z}{2 \cdot \sqrt{D_{geo,eff} \cdot t}}\right)\right) \qquad (1)$$

Where $c_{geo}$ is the *Geobacter* cell density in the pore water (cells·L$^{-1}$), $c_{geo,top}$ is the *Geobacter* cell density in the top layer (cells·L$^{-1}$), $t$ is the time (day), $z$ is the distance from the top layer (cm) and $D_{geo,eff}$ (cm$^2$·day$^{-1}$) is the effective random motility coefficient for *Geobacter* in the sand. Optimization was performed using least square difference minimization with Microsoft Excel Solver and the macro SolverAid of Macrobundle was used to calculate the standard error. As we explained previously (Philips et al., 2011), the porosity of the sand $\theta$ can be used to calculate the tortuosity $\xi$ of the sand, which relates the effective random motility coefficient in sand to the random motility coefficient in water ($D_{geo,0}$):

$$D_{geo,eff} = \xi \cdot D_{geo,0} = \theta^{1/2} \cdot D_{geo,0} \qquad (2)$$

## 2.7 Modeling

The results of the diffusion-cell experiment were analyzed with a diffusion-motility model to clarify how the motility of *Geobacter* contributed to the dechlorination. We previously used a model that incorporated TCE self-inhibition, i.e. inhibition of the dechlorination reaction at elevated TCE concentrations, to describe TCE dechlorination and growth of *Geobacter* in liquid batch systems (Philips et al., 2013a). That existing model was here extended to a diffusion-motility model that included a random motility term for *Geobacter,* analogue to the diffusion term for TCE. Chemotaxis was not incorporated, since no experimental proof for chemotaxis towards or away of TCE was found. The exponential growth term used previously was replaced with a logistic growth term, for reasons explained in the results section. This logistic growth term assumes that the growth rate declines if a maximum *Geobacter* cell density is approached (McMeekin et al., 1993). The resulting diffusion-motility model is described by the system of partial differential equations:

$$\begin{cases} \dfrac{\partial c_{TCE}}{\partial t} = D_{TCE} \cdot \dfrac{\partial^2 c_{TCE}}{\partial z^2} - k_{cell} \cdot c_{GEO} \\ \dfrac{\partial c_{DCE}}{\partial t} = D_{DCE} \cdot \dfrac{\partial^2 c_{DCE}}{\partial z^2} + k_{cell} \cdot c_{GEO} \\ \dfrac{\partial c_{GEO}}{\partial t} = D_{GEO} \cdot \dfrac{\partial^2 c_{GEO}}{\partial z^2} + Y_{GEO} \cdot k_{cell} \cdot c_{GEO} \cdot \left(1 - \dfrac{c_{GEO}}{c_{GEO,\max}}\right) - k_d \cdot c_{GEO} \end{cases} \quad (3)$$

Where $c_{TCE}$ is the TCE concentration (mmol·L$^{-1}$), $c_{DCE}$ is the DCE concentration (mmol·L$^{-1}$), $c_{GEO}$ is the *Geobacter* cell density (cells·L$^{-1}$), $t$ is the time (day), $z$ is the distance from the DNAPL layer (cm), $Y_{GEO}$ is the growth yield of *Geobacter* (cells·mmol$^{-1}$), $k_d$ is the microbial decay coefficient (day$^{-1}$), $c_{GEO,max}$ is the maximum *Geobacter* cell density (cells·L$^{-1}$), $D_{TCE}$ is the diffusion coefficient for TCE (cm$^2$·day$^{-1}$), $D_{DCE}$ is the diffusion coefficient for DCE (cm$^2$·day$^{-1}$), and $D_{GEO}$ is the random motility coefficient for *Geobacter* (cm$^2$·day$^{-1}$). The cell specific dechlorination rate $k_{cell}$ (mmol·cell$^{-1}$·day$^{-1}$) depends on the TCE concentration (Haest et al., 2010a):

$$k_{cell} = \dfrac{k_{cell,\max} \cdot c_{TCE}}{(K_{s,TCE} + c_{TCE}) \cdot \left(1 + \left(\dfrac{c_{TCE}}{EC_{50}}\right)^{b_i}\right)} \quad (4)$$

Where $k_{cell,max}$ is the maximum cell specific dechlorination rate (mmol·cell$^{-1}$·day$^{-1}$), $K_{s,TCE}$ is the half saturation concentration (mmol·L$^{-1}$), $EC_{50}$ is the TCE concentration at which the cell specific dechlorination rate is halved compared to the maximum rate because of the TCE self-inhibition (mmol·L$^{-1}$) and $b_i$ (-) is related to the slope of the cell specific dechlorination rate at the $EC_{50}$ concentration.

Retardation of TCE in the central layer was not incorporated, since the used sand was inert and contained no organic matter. In addition, we assumed that all *Geobacter* cells remained in the pore water and did not attach to the sand.

The system of partial differential equations (3) was solved over the domain $z \in [-3.5, 8.5]$. This domain was split into three zones to model the three-layers of the diffusion-cells, setting the domain for the bottom, central and top layer respectively [-3.5,0], [0,5.5] and [5.5,8.5] (Figure S1). For the central layer, a porosity $\theta$ was assumed. For the top layer, the diffusion and random motility coefficients in water were used, while for the central layer these coefficients were adjusted using the tortuosity as described by Equation (2) (Figure S1A). For the bottom layer, only the *Geobacter* cell densities were modeled, the TCE and DCE concentrations were set constant in this layer (Figure S1D). The effect of the mixing in the DNAPL layer on the *Geobacter* cell densities was modeled by setting the random motility coefficient thousand times higher than that in water (Figure S1A). The diffusion and motility fluxes at both sides of the boundary between different layers were set equal, manifesting the conservation of mass and biomass respectively (Figure S1A). The initial conditions and boundary conditions used in the model are depicted in Figure S1B-D.

The system of reaction-diffusion equations (3) was solved using an explicit-implicit method. We employed the Crank-Nicholson scheme to handle the linear terms in system (3), while for the nonlinear terms we used the predictor-corrector method in order to avoid solving nonlinear systems of algebraic equations (Thomas, 1995; Butcher, 2003). All parameters were derived from experiments described here or previously (Table S1), except for the parameter $k_{cell,max}$, which was adjusted to fit the experimental results (described below).

The sensitivity of the diffusion-motility model towards its various parameters was analyzed using Morris' one-at-a-time (OAT) design (Morris, 1991; Mertens et al., 2005). Shortly, this method determines the distribution of the elementary effect each parameter has on the model outcome using different parameter combinations. We evaluated a total of 1000 parameter combinations. The effect of $D_{GEO}$, $k_{cell,max}$, $EC_{50}$, $b_i$, $K_{s,TCE}$, $Y_{GEO}$, $k_d$ and $c_{GEO,max}$ was investigated over the parameter intervals given in Table S1, which are mostly chosen as a factor 3 around the parameter values used by the model. Other model parameters were assumed constant. The effect on the model outcome was evaluated as the root mean square error between the modelled and experimental TCE and *cis*-DCE concentrations at day 19 with both sand and top layer inoculation and for all three different inoculation densities.

## 3 RESULTS

### 3.1 Chemotaxis assay

The agarose-in-plug bridge method (Yu and Alam, 1997) was used to test for chemotaxis towards or away from TCE. Microscopic observation of the bacterial cells around the plugs showed that several cells were motile. Nevertheless, the distribution of cells around the plugs with TCE was similar to that around control plugs (Figure 1). No accumulation of cells at or away from the edge of the TCE-containing plugs could be seen, as has previously been observed for other bacteria using similar plugs containing TCE (Parales et al., 2000; Singh and Olson, 2010). As such, no evidence for chemotaxis by the KB-1™ subculture towards or away from TCE was found.

### 3.2 Motility experiment

The motility of the KB-1$^{TM}$ subculture was studied by inoculating an aqueous layer on top of a sterile sand layer. Measured 16S rRNA gene copy numbers at day 4 were generally lower than at day 7 at corresponding distances from the top layer (Figure S2). By day 7, the *Geobacter* cells had migrated over a distance of at least 4.4 cm (Figure 2). At this distance, the *Geobacter* cell density was two orders of magnitude lower than in the top layer. Over the same distance, bacterial 16S rRNA gene copy numbers dropped only one order of magnitude. Also the RFLP profiles suggested that the *Geobacter* subpopulation had migrated slower than the total bacterial population, as only the RFLP profiles for the two samples closest to the top layer showed correspondence with the RFLP profile of *Geobacter lovleyi* SZ (Figure 2). In the middle of the examined sand layer part, RFLP profiles corresponded with the RFLP profiles of the Clostridium1 and Clostridium7 operational taxonomic units (OTU) (Philips et al., 2012). The RFLP profiles for the samples at the largest distance from the top layer corresponded with the RFLP profile of a yet unidentified species (Philips et al., 2012). The faster migration of the fermentative species in comparison to *Geobacter* could be explained by a higher motility, but is in part likely also due to their growth on the yeast extract and cysteine in the medium (Philips et al., 2013c).

The *Geobacter* 16S rRNA gene copy numbers measured in this experiment were used to calculate a random motility coefficient for *Geobacter*. The use of Equation 1 presumes no growth for *Geobacter*, which is likely correct since no TCE was present in the systems. Equation 1 fitted the experimental data rather well, as a coefficient of determination ($R^2$) of 0.79 was obtained between log transformed observed and predicted cell densities (Figure S2). This resulted in an effective random motility coefficient in the sand of 0.15 ± 0.02 cm$^2$ day$^{-1}$, which corresponds to a random motility coefficient in water of 0.24 ± 0.02 cm$^2$ day$^{-1}$. This value is well in the range of previously reported motility coefficients (Ford and Harvey, 2007) and is of the same order of magnitude as the diffusion coefficient of TCE (Table S1).

### 3.3 Diffusion-cell experiment

Diffusion-cells were used as model source zones to investigate the effect of dechlorinator motility on TCE DNAPL dissolution. To allow comparison with previous experiments, a first set of diffusion-cells was inoculated in the central sand layer. Three different inoculation densities were applied, but did not result in significantly different TCE and *cis*-DCE concentration profiles 19 days after inoculation (Figure 3A). In all diffusion-cells with sand layer inoculation, the TCE concentration at day 19 declined from the saturated TCE concentration (8.4 mM) at the DNAPL layer to a concentration below 0.6 mM at a distance of 2.0 to 2.5 cm from the DNAPL (Figure 3A). The *cis*-DCE concentration was maximal at about 2 cm distance from the DNAPL and decreased towards the DNAPL and the top layer. These concentration profiles resulted in a DNAPL dissolution enhancement of a factor 2.5 ± 0.3 in comparison to abiotic dissolution (calculated as described by Philips et al. (2011)). We previously showed that similar TCE and *cis*-DCE concentration profiles remained unchanged beyond day 18 (Philips et al., 2011). As such, with sand layer inoculation, the TCE and *cis*-DCE concentration profiles measured at day 19 were in steady-state, irrespectively of the applied inoculation density.

A second set of diffusion-cells was inoculated in the top layer to investigate the effect of dechlorinator motility on TCE DNAPL dissolution. Similarly as for the first set of diffusion-cells, concentration profiles in the diffusion-cells with top layer inoculation were independent of the applied inoculation density (Figure 3B). Before inoculation of the top layer, the linearly decreasing abiotic steady-state TCE concentration profile was established as described in the methods section. Nineteen days later, the TCE concentration in all diffusion-cells with top layer inoculation declined below the detection limit at about 3.5 cm from the DNAPL, whereas the *cis*-DCE concentration was maximal at about 2.5 cm distance from the DNAPL (Figure 3B). This resulted in a DNAPL dissolution enhancement factor of 1.9 ± 0.3 in comparison to abiotic dissolution. These results demonstrate that *Geobacter* cells, migrating out of the top layer, were able to dechlorinate in the sand layer and succeeded in enhancing the initial DNAPL dissolution flux. At day 19, however, the TCE concentration profiles were 24% less steep than in the first set of diffusion-cells (Figure 3). Pore water sampling of the second set of diffusion-cells was therefore repeated at day 26 and by that time, the TCE concentration profiles were similar (results not shown) to those measured at day 19 in the diffusion-cells with sand layer inoculation (Figure 3A). This shows that with top layer inoculation, the concentration profiles were not yet at steady-state at day 19, whereas steady-state concentrations were attained one week later.

An additional diffusion-cell in each set was sampled to examine the microbial distribution. Pore water samples taken at day 19 from the diffusion-cell with sand layer inoculation showed two distinctive zones (Figure 4A). In the lower part of the sand layer, *Geobacter* 16S rRNA gene copy numbers were high and comparable to the bacterial 16S rRNA gene copy numbers. Even adjacent to the DNAPL layer, where the TCE concentrations were toxic (> 2.6 mM), *Geobacter* cell densities were almost two orders of magnitude higher than at inoculation. The RFLP profiles for the lower part of the sand layer corresponded with the RFLP profile of *Geobacter lovleyi* SZ. In the upper part of the sand layer, *Geobacter* 16S rRNA gene copy numbers were one order of magnitude lower than bacterial 16S rRNA gene copy numbers. The RFLP profiles for this part of the sand layer corresponded with the RFLP of an unidentified species (Philips et al., 2012). Similar results were previously related to the TCE concentration (Philips et al., 2012).

For the diffusion-cell with top layer inoculation, the pore water samples taken at day 19 showed a microbial distribution consisting of three zones (Figure 4B). In the upper sand layer part of this diffusion-cell, *Geobacter* and bacterial 16S rRNA gene copy numbers and RFLP profiles were similar to those in the upper sand layer part of the diffusion-cell with sand layer inoculation (Figure 4). In the middle part of the sand layer, where the TCE dechlorination occurred, the *Geobacter* 16S rRNA gene copy numbers were high and similar to the bacterial 16S rRNA gene copy numbers (Figure 4B). The RFLP profiles for this part of the sand layer corresponded with the RFLP profile of *Geobacter lovleyi* SZ. Closer to the DNAPL layer, both *Geobacter* and bacterial cell densities dropped several orders of magnitude and the RFLP profiles had a low intensity (Figure S2B). These results suggest that by day 19 the *Geobacter* cells migrating out of the top layer had not yet arrived in the zone adjacent to the TCE DNAPL.

### 3.4 Modeling results

A diffusion-motility model (Equation 3) was used to analyze the results of the diffusion-cell experiment. This model used the kinetic and growth parameters previously derived from a liquid batch experiment (Philips et al., 2013a), while the random motility coefficient for *Geobacter* ($D_{GEO}$) was obtained from the motility experiment of this study (Table S1). Initially, the diffusion-motility model was solved assuming exponential growth for *Geobacter,* similarly as in our previous batch model (Philips et al., 2013a). This initial model, however, predicted a maximum *Geobacter* cell density of $1 \cdot 10^9$ cells·mL$^{-1}$ (results not shown), whereas in the diffusion-cells the *Geobacter* cell densities were not higher than $5 \cdot 10^7$ copies·mL$^{-1}$ (Figure 4). As a consequence, the TCE dechlorination rate was overestimated, as this model predicted a complete conversion of TCE to *cis*-DCE at less than 1 cm distance from the DNAPL layer (results not shown). Experimental *Geobacter* cell densities in our previous batch experiment were also not higher than $5 \cdot 10^7$ copies·mL$^{-1}$ (Philips et al., 2013a), but exponential growth described the growth of *Geobacter* well, likely since in the batch systems *Geobacter* only grew on a single addition of TCE, resulting in a limited increase in cell numbers. With a continuous TCE supply as in the diffusion-cells, however, chemical and physical limitations are likely to constrain growth beyond a certain maximum cell density. For this reason, logistic growth was incorporated in the diffusion-motility model (Equation 3) and the maximum *Geobacter* cell density observed in the diffusion-cell experiment was set as $c_{GEO,max}$ (Table S1). Similarly, Haest et al. (2010b) incorporated a reduced dechlorination rate at elevated dechlorinator cell densities in order to describe dechlorination in flow-through columns.

The diffusion-motility model incorporating logistic growth and using the kinetic parameters as derived from a previous batch experiment (Philips et al., 2013a) underestimated the dechlorination rate. At day 19, the model predicted that steady-state chlorinated ethene concentration profiles would only be attained for the treatment combining sand layer inoculation with a high inoculation dose (Figure S3), while the steady-state TCE concentration profile was somewhat less steep than experimentally observed. For this reason, the parameter value for $k_{cell,max}$ was increased with a factor of 1.9 in comparison to the previously derived value (Philips et al., 2013a) to fit the chlorinated ethene concentration profiles. The maximum cell specific dechlorination rate ($k_{cell,max}$) is highly correlated with the *Geobacter* growth yield, which has an uncertainty of more than a factor two due to the variability in the quantification of microbial numbers by qPCR (Philips et al., 2012). This uncertainty could explain why a doubling of $k_{cell,max}$ was required to describe dechlorination in the diffusion-cells. Alternatively, the required increase of $k_{cell,max}$ might be due to a higher maximum cell specific dechlorination rate in the sand then in batch, as has previously been reported by others. Schaefer et al. (2009), for instance, explained that only a 200-times higher relative dechlorination rate in comparison to batch could describe dechlorination in their soil column experiment. In addition, Sabalowsky and Semprini (2010) had to adjust their kinetic parameters obtained from batch in order to describe dechlorination in a continuous flow stirred tank reactor and a recirculating packed column.

The diffusion-motility model using the adjusted parameter value for $k_{cell,max}$ fitted the measured chlorinated ethene concentration profiles well (Figure 5). For sand layer inoculation, this model correctly described that steady-state chlorinated ethene concentration profiles were attained before day 19 at all inoculation densities (Figure 5A). In addition, for top layer inoculation, the model correctly predicted that chlorinated ethene concentration profiles were not yet at steady-state by day 19 (Figure 5B). The modeled transient TCE concentration profiles, however, varied with different inoculation densities (Figure 5B), in contrast to the experimental results (Figure 3B). Furthermore, the model correctly described the high *Geobacter* cell densities close the to DNAPL with inoculation of the sand layer (Figure 5A), while for top layer inoculation the low *Geobacter* cell densities in the lower part of the sand layer were overestimated by the model. In addition, the model overestimated the *Geobacter* cell densities in the upper part of the sand layer with both inoculation methods (Figure 5).

A Morris' OAT sensitivity analysis was performed to investigate the sensitivity of the model towards its various parameters (Figure S4). This sensitivity analysis shows that the random motility coefficient of *Geobacter* has an intermediate effect on the model outcome, while the model was most sensitive to the growth parameters $Y_{GEO}$ and $c_{GEO,max}$ and to the kinetic parameters $k_{cell,max}$ and $EC_{50}$. The parameters $b_i$, $K_{s,TCE}$ and $k_d$ had a low impact on the model, which is in agreement with our previous findings (Haest et al., 2010b; Philips et al., 2013a). Visualization of the sensitivity of the model towards some of its parameters shows that a factor 4 change in $D_{GEO,0}$ has a low effect on the modeled TCE concentrations, but has a clear impact on the modeled *Geobacter* cell densities (Figure S4A). In addition, a factor 4 difference of the growth yield $Y_{geo}$ shows a large effect on both the modeled TCE concentrations and *Geobacter* cell densities (Figure S4B).

The diffusion-motility model allows to investigate the case of immotile *Geobacter* dechlorinators. Immotile colloids of the size of a bacterium have a Brownian diffusion coefficient which is three orders of magnitude lower than the random motility coefficient for *Geobacter* reported in this study (Ford and Harvey, 2007). As such, the case of immotile *Geobacter* cells was modeled by setting the value for $D_{GEO,0}$ $10^3$ times lower than reported in Table S1. For sand layer inoculation, immotile *Geobacter* cells would not have resulted in different chlorinated ethene concentration profiles than experimentally observed, but *Geobacter* cell densities would only have been high in the middle part of the sand layer where dechlorination occurs (Figure S6A). For top layer inoculation, however, the model shows that by day 19 immotile *Geobacter* cells would have migrated only a few millimeters in the sand layer, while there would have been hardly any enhancement of the DNAPL dissolution (Figure S6B).

# 4 DISCUSSION

## 4.1 The combination of random motility and growth explains the rapid enhancement of TCE DNAPL dissolution in diffusion-cells

The dechlorinators in the studied KB-1$^{TM}$ subculture are related to *Geobacter lovleyi* SZ (Philips et al., 2012), for which flagellar motility was previously described (Sung et al., 2006). Microscopic observation confirmed that the cells in the studied KB-1$^{TM}$ subculture were motile, but no evidence for chemotaxis towards or away from TCE was found (Figure 1). Nevertheless, chemotaxis genes and associated genes expressing yet unknown signaling functions have been found in the genome of several *Geobacter* species (Tran et al., 2008). As such, a further investigation of chemotaxis by the *Geobacter* dechlorinators in the KB-1$^{TM}$ culture towards other potential attractants and repellents, like electron donors, nutrients and degradation products, is warranted. In addition, it was found that the fermentative *Clostridium* species in the KB-1$^{TM}$ subculture were motile and even migrated faster than the *Geobacter* dechlorinators (Figure 2). It should further be examined whether bioaugmentation can benefit for the motility of these fermenters. This study, however, continued by investigating whether the random motility of the *Geobacter* dechlorinators could contribute to an effective bioaugmentation of TCE DNAPL source zones.

A motility experiment, in which an inoculated top layer was placed above a sterile sand layer, showed that the *Geobacter* dechlorinators were able to migrate several centimeters into the sand layer in just 7 days (Figure 2). In addition, the migration of the *Geobacter* dechlorinators was examined in model source zones using a diffusion-cell setup. The diffusion-cell experiment demonstrated that *Geobacter* cells migrating out of the top layer managed to rapidly enhance the initial TCE DNAPL dissolution flux in the sand layer (Figure 3B). The developed diffusion-motility model (Equation 3) allowed to simulate the case of immotile *Geobacter* dechlorinators, which suggested the important role of random motility in the fast colonization of the sand layer (Figure S6B). The sensitivity analysis, however, found that the random motility coefficient only had an intermediate impact on the modeled chlorinated ethene concentrations and that the model was most sensitive to the growth and kinetic parameters (Figure S4). Consequently, it was the combined effect of random motility and growth of the *Geobacter* dechlorinators on the dechlorination of TCE that explains the rapid DNAPL dissolution enhancement in the diffusion-cells.

### 4.2 The experimental *Geobacter* cell densities can be explained by random motility

In the diffusion-cell with sand layer inoculation, *Geobacter* cell densities adjacent to the DNAPL layer were significantly higher than at inoculation, even though the TCE concentrations in this zone were toxic (> 2.6 mM) (Figure 4A). We obtained similar findings from previous diffusion-cell experiments and argued before that these results might be explained by random motility (Philips et al., 2012). The diffusion-motility model developed in this study well described the high *Geobacter* cell densities adjacent to the DNAPL (Figure 5A). Simulating immotile *Geobacter* cells, this model showed that the *Geobacter* cell densities would remain close to the inoculation density in the lower part of the sand layer and would increase almost two orders of magnitude around the distance from the DNAPL where the TCE concentration drops below 2.6 mM (Figure S6A). As such, the diffusion-motility model indicates that random motility was the main mechanism for high *Geobacter* cell densities close to the DNAPL in the diffusion-cell with sand layer inoculation.

The diffusion-motility model overestimated the low *Geobacter* cell densities measured adjacent to the DNAPL in the diffusion-cell with top layer inoculation (Figure 5B). However, the diffusion-motility model simulated the experimental *Geobacter* cell densities better by decreasing the random motility coefficient for *Geobacter* ($D_{GEO,0}$) with a factor 4 (Figure S5A). The parameter $D_{GEO,0}$ was fit on *Geobacter* 16S rRNA gene copy numbers quantified by qPCR analysis, for which we previously found a large variability (Philips et al., 2012). As such, the clear overestimation of low *Geobacter* cell densities measured adjacent to the DNAPL with top layer inoculation could be due to an inaccurate random motility coefficient.

Experimental *Geobacter* cell densities in the upper part of the sand layer were significantly lower than in the dechlorination zone with both inoculation methods (Figure 4). In previous diffusion-cell experiments, we observed a similar drop of the *Geobacter* cell densities of about one order of magnitude between the zones of the sand layer with and without TCE (Philips et al., 2012; Philips et al., 2013c). The diffusion-motility model, however, overestimated the *Geobacter* cell densities in the upper sand layer part (Figure 5). By assuming random motility, the diffusion-motility model predicts the development of a constant *Geobacter* cell density equal to $c_{GEO,max}$ throughout most of the sand layer (Figure 5A: HD). As such, the low *Geobacter* cell densities in the upper part of the sand layer are in contradiction with random motility and rather suggest that the *Geobacter* dechlorinators have some sort of mechanism to accumulate in the lower part of the sand layer where TCE is present. Nevertheless, random motility was sufficient to describe the TCE dechlorination patterns throughout the diffusion-cells and the *Geobacter* cell densities in the other parts of the sand layer (Figure 5).

### 4.3 Low impact of the inoculation density

The inoculation density did not affect the chlorinated ethene concentration profiles measured in the different diffusion-cells (Figure 3). Similar chlorinated ethene concentration profiles were even recorded for top layer inoculation, although the concentration profiles were still transient at the time of sampling (Figure 3). The diffusion-motility model, in contrast, predicted that transient concentration profiles would depend on the inoculation density (Figure 5B). However, both the experimental and model results agreed that steady-state chlorinated ethene concentration profiles were independent of the inoculation density (Figure 5), illustrating the low impact of the applied inoculation density on the dechlorination in the diffusion-cells. Similarly, Haest et al.(2010b) found by modeling that the initial number of dechlorinators was of low importance for dechlorination in flow-through columns. In the diffusion-cells of the current study, the different inoculation densities led to initial differences in cell densities and dechlorination rates, but these differences were likely readily eliminated due to a fast microbial growth.

### 4.4 Implications for source zone bioremediation

Bioaugmentation is seen as a valuable strategy to initiate dechlorination in TCE contaminated aquifers (Major et al., 2002; Adamson et al., 2003; Sleep et al., 2006). In a contaminant plume, the ground water flow can help spreading an applied inoculum, but a DNAPL is typically formed in a low-permeability zone and is thus difficult to reach via the ground water. Therefore, the success of bioaugmentation of a DNAPL source zone largely depends on the migration mechanisms of the applied bacteria, (Singh and Olson, 2008). The *Geobacter* dechlorinators in the studied KB-1$^{TM}$ subculture are not chemotactic towards or away from TCE, but their random motility resulted in a fast colonization of the sand layer and the biological enhancement of TCE DNAPL dissolution. As such, *Geobacter* containing dechlorinating cultures seem a suitable choice for the bioaugmentation of TCE DNAPL source zones. It should be noted, however, that the diffusion-cell setup used in this study only simulates a simplified DNAPL source zone, in which no bacteria can exit the system or migrate out the source zone. At the edge of a real DNAPL source zone, the groundwater flow likely rinses a large part of the applied bacteria away. Bacteria showing positive chemotaxis towards the contaminant, therefore, might be more effective in colonizing a DNAPL source zone that just randomly motile bacteria. Using the diffusion-cell setup, we demonstrated that different inoculation levels had no effect on the final dechlorination rate. In real source zones, however, the number of bacteria that overcome the groundwater flow and effectively enter the source zone might largely depend on the applied inoculation dose. Finally, it should be noted that this study only investigated the degradation of TCE to *cis*-DCE, since dechlorination beyond *cis*-DCE is mostly inhibited in DNAPL source zones (Yang and McCarty, 2000; Adamson et al., 2004). Only the stimulation of *cis*-DCE to ethene dechlorination downstream of the DNAPL source zone will allow for a complete decontamination of TCE.

**ACKNOWLEDGEMENTS**

We thank Fanny Hamels for her assistance in performing the experimental work. This research was funded as a Ph.D. Fellowship of the Research Foundation – Flanders (FWO).

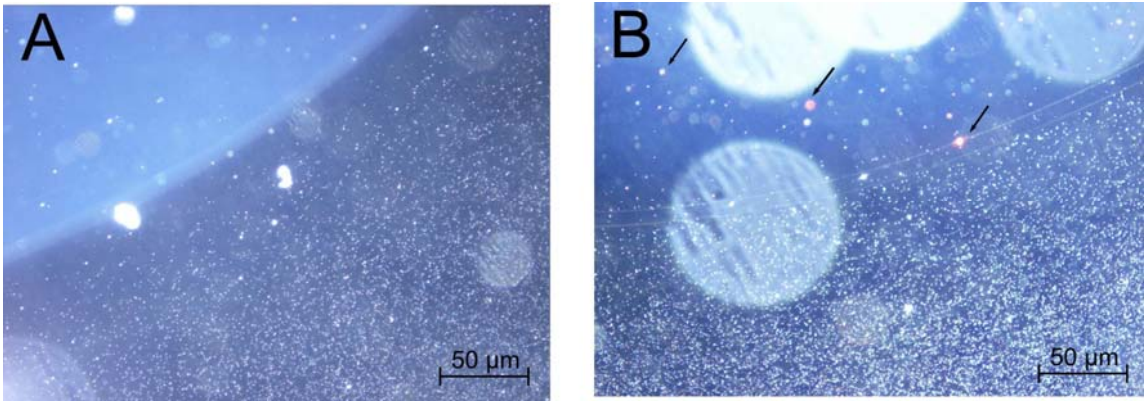

Figure 1: Chemotaxis assay using agarose plugs without TCE (A) and with TCE (stained with Oil-Red-O) (B). The TCE droplets in the agarose plug of B can be seen by their red color (indicated by arrows). The small white dots are individual cells.

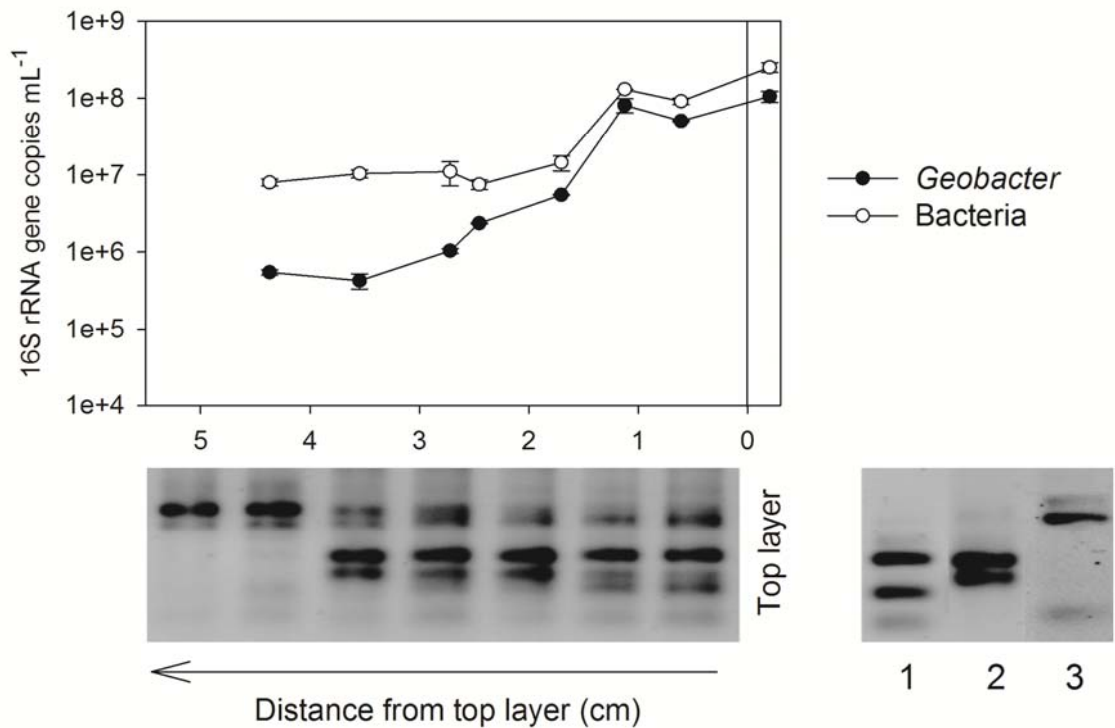

Figure 2: Number of *Geobacter* and bacterial 16S rRNA gene copies per mL pore water and RFLP profiles in relation to the distance from the top layer for samples taken seven days after inoculation of an aqueous layer on top of a sterile sand layer. Top layers were inoculated with respectively $8 \cdot 10^7$ and $2 \cdot 10^8$ *Geobacter* and bacterial 16S rRNA gene copies per mL. The RFLP profiles are given in the order of the corresponding 16S rRNA gene copy number data points of the graph above. Error bars on the 16S rRNA gene copy numbers indicate the standard deviation on duplicate qPCR measurements of the same extracted DNA sample. Numbered control lanes correspond as follows: 1, *Geobacter lovleyi* SZ; 2, Clostridium1 OTU; and 3, Clostridium7 OTU (Philips et al., 2012).

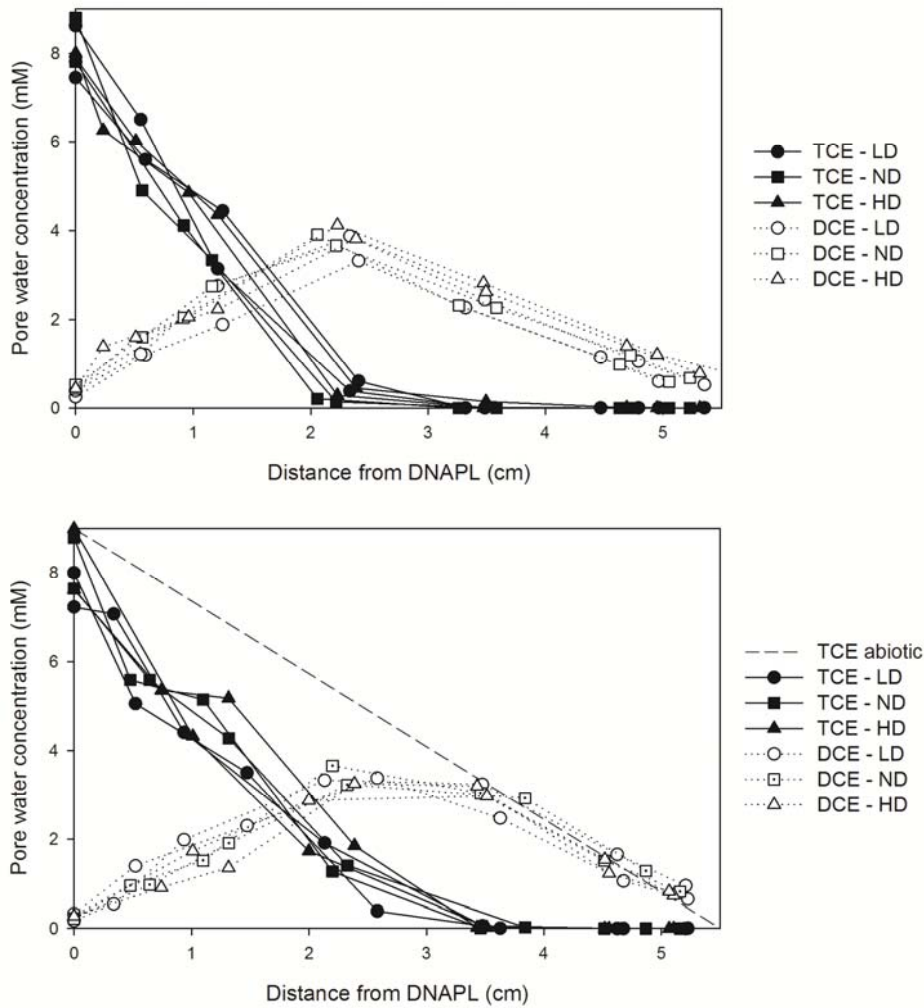

Figure 3: Pore water TCE and *cis*-DCE concentrations in relation to the distance from the DNAPL layer for samples taken from diffusion-cells 19 days after inoculation of the sand layer (A) or of the top layer (B) using three different inoculation densities, i.e. LD: $4 \cdot 10^5$, ND: $4 \cdot 10^6$ and HD: $4 \cdot 10^7$ *Geobacter* 16S rRNA gene copies per mL. Before inoculation of the top layer, the linearly decreasing abiotic steady-state TCE concentration profile (dashed line) was established. Different inoculation densities did not result in significantly different TCE and *cis*-DCE concentration profiles.

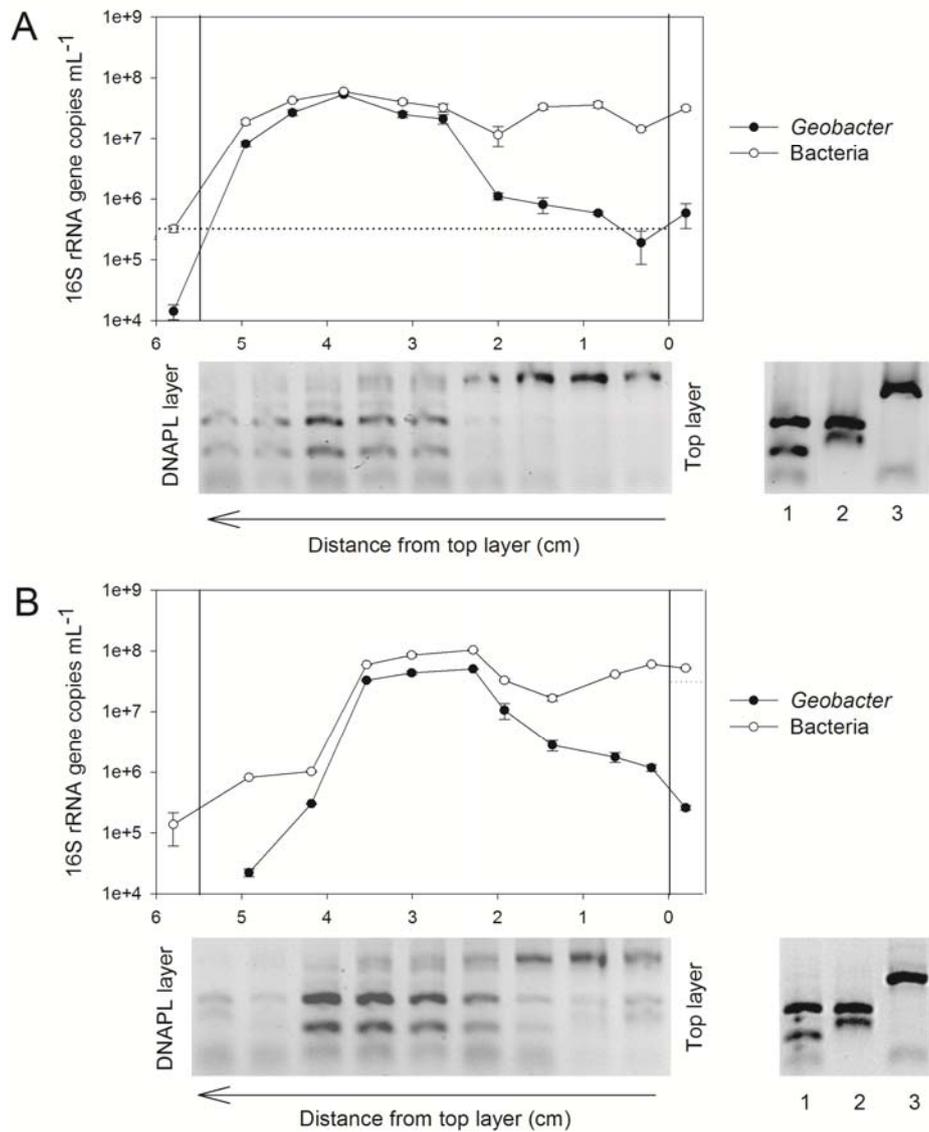

Figure 4: The number of *Geobacter* and bacterial 16S rRNA gene copies per mL pore water and RFLP profiles in relation to the distance from the top layer for diffusion-cells sampled 19 days after inoculation of the sand layer (A) or of the top layer (B). Before inoculation of the top layer, the linearly decreasing abiotic steady-state TCE concentration profile was established. The initial *Geobacter* copy numbers are indicated by the dotted lines. The RFLP are given in the order of the corresponding 16S rRNA gene copy number data points of the graph above. Error bars on the 16S rRNA gene copy numbers indicate the standard deviation on duplicate qPCR measurements of the same extracted DNA sample. Numbered control lanes correspond as follows: 1, *Geobacter lovleyi* SZ; 2, Clostridium1 OTU; and 3, Clostridium7 OTU (Philips et al., 2012).

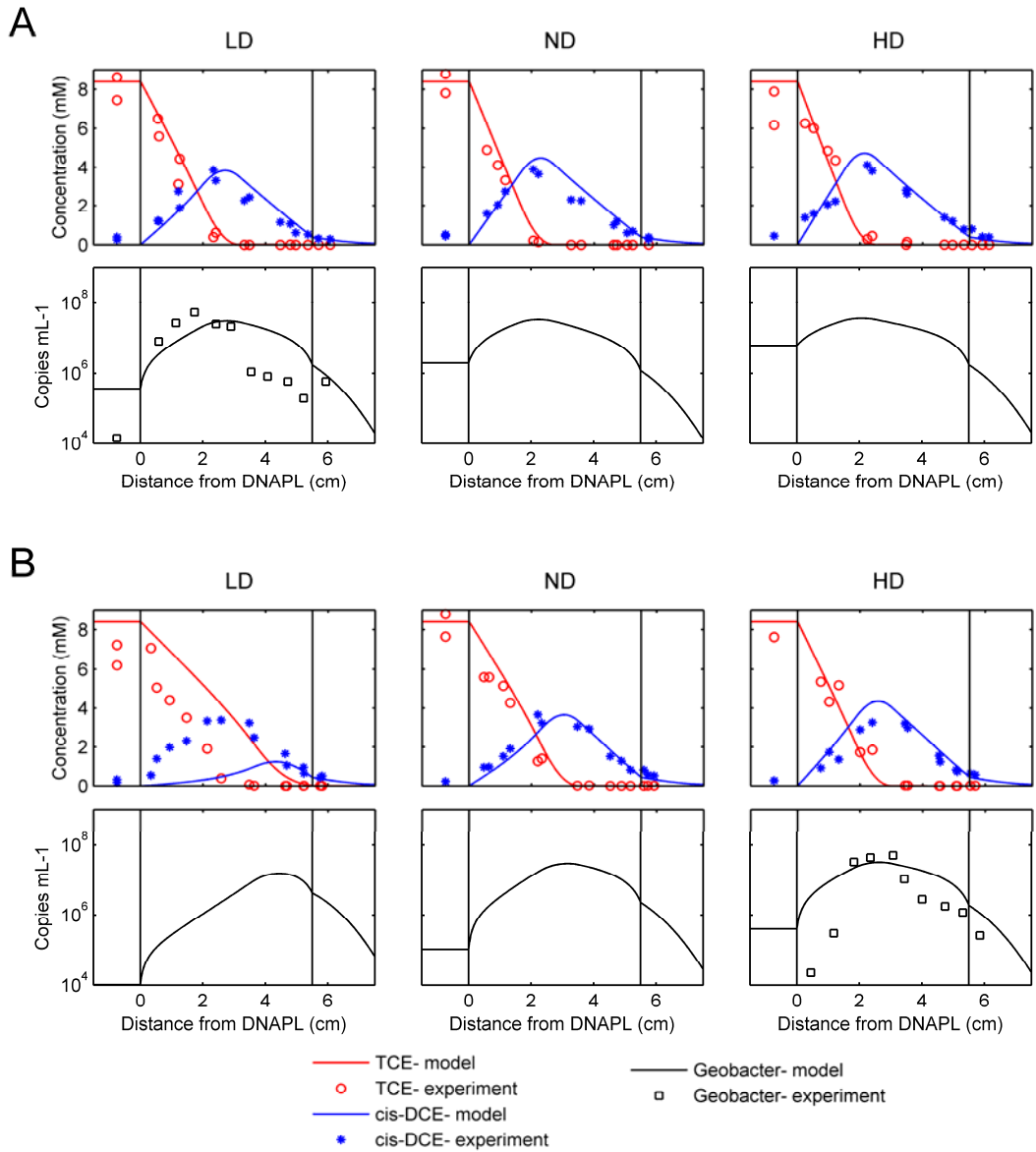

Figure 5: Modeled TCE and *cis*-DCE concentrations and *Geobacter* 16S rRNA gene copy numbers according to the diffusion-motility model versus the experimental results obtained at day 19 after inoculation of the sand layer (A) or the top layer (B) of diffusion-cells, using three different inoculation densities, i.e. LD: $4·10^5$, ND: $4·10^6$ and HD: $4·10^7$ *Geobacter* 16S rRNA gene copies per mL. Experimental *Geobacter* 16S rRNA gene copy numbers were only determined for sand layer inoculation combined

with a low inoculation density and for top layer inoculation combined with a high inoculation density.

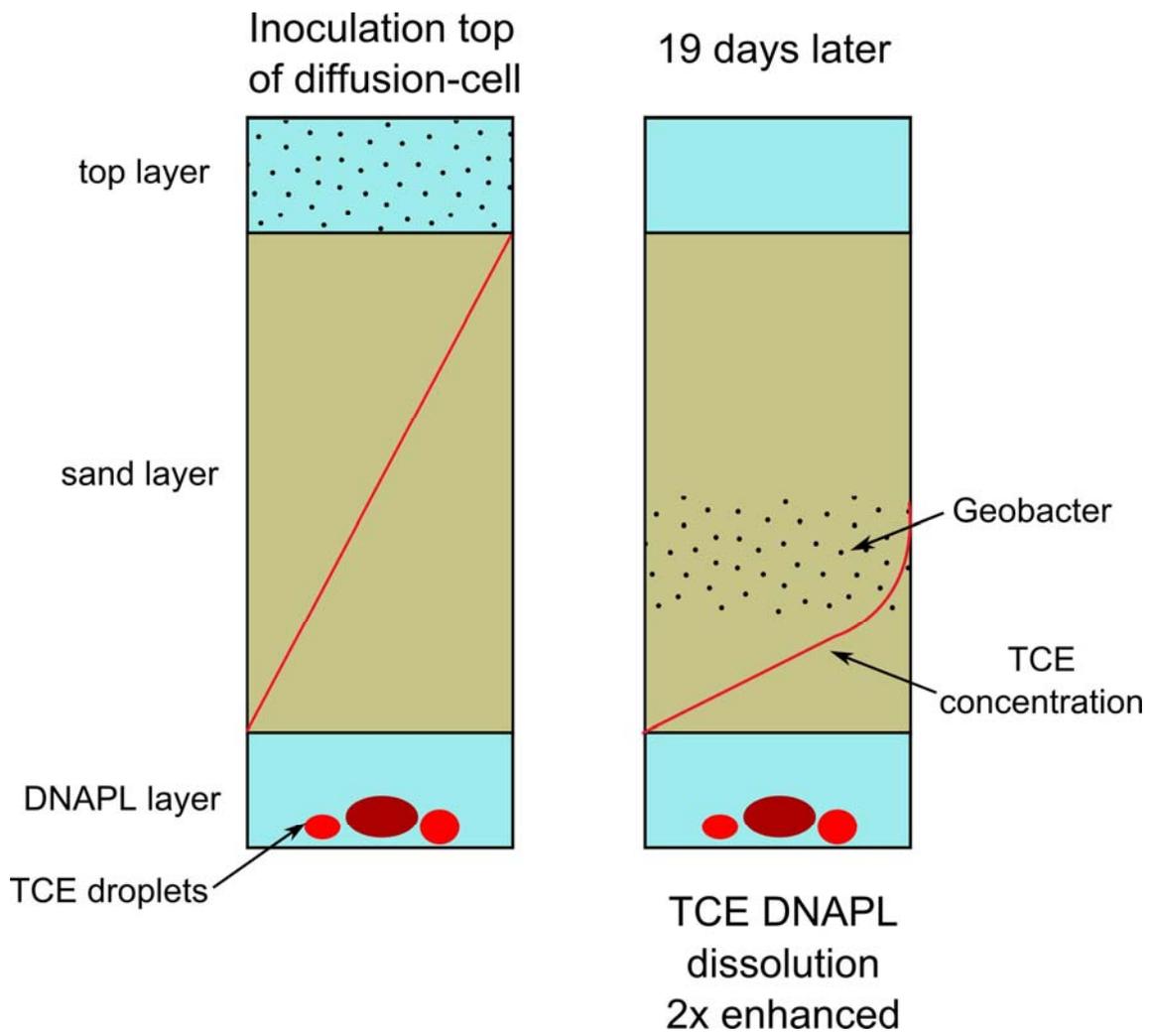

Graphical abstract

# SUPPORTING INFORMATION:
# Motile *Geobacter* dechlorinators migrate into a model source zone of trichloroethene dense non-aqueous phase liquid: experimental evaluation and modeling

Jo Philips [a,1,*] (jo.philips@ugent.be)
Alexey Miroshnikov [b] (alexeym@cns.umass.edu)
Pieter Jan Haest [a,2] (pieterjan.haest@agt.be)
Dirk Springael [a] (dirk.springael@ees.kuleuven.be)
& Erik Smolders [a] (erik.smolders@ees.kuleuven.be)

[a]  Department of Earth and Environmental Sciences, Division of Soil and Water Management, KU Leuven (University of Leuven),
Kasteelpark Arenberg 20, B-3001 Heverlee, BELGIUM
Tel: +3216321609, Fax: +3216321997

[b]  Department of Mathematics and Statistics, University of Massachusetts,
Lederle Graduate Research Tower, Amherst MA 01003-9305, USA

*  Corresponding author

[1]  Current address: Laboratory of Microbial Ecology and Technology, Ghent University, Coupure Links 653A, B-9000 Ghent, BELGIUM

[2]  Current address: A.G.T. n.v., Kontichsesteenweg 38, B-2630 Aartselaar, BELGIUM

Supporting Information includes:

   9 pages

   1 table

   5 figures

Table S1: Parameter values used in the diffusion-motility model and intervals used for the sensitivity analysis.

| Parameter | Value | Interval sensitivity analysis[f] |
|---|---|---|
| $k_{cell,max}$ (mmol·cell$^{-1}$·day$^{-1}$) | $6.0 \cdot 10^{-11}$ [a] | $1.0 \cdot 10^{-11}$ -- $9.0 \cdot 10^{-11}$ |
| $EC_{50}$ (mmol·L$^{-1}$) | 2.13 [b] | 1 - 4 |
| $b_i$ (-) | 12.66 [b] | 8 - 14 |
| $K_{s,TCE}$ (mmol·L$^{-1}$) | $4.19 \cdot 10^{-3}$ [c] | $1.4 \cdot 10^{-3} - 1.3 \cdot 10^{-2}$ |
| $Y_{GEO}$ (cell·mmol$^{-1}$) | $9.4 \cdot 10^{9}$ [b] | $3 \cdot 10^{9} - 3 \cdot 10^{10}$ |
| $k_d$ (day$^{-1}$) | 0.031 [b] | 0.01 – 0.09 |
| $c_{GEO,max}$ (cell·mL$^{-1}$) | $5 \cdot 10^{7}$ [e] | $1.5 \cdot 10^{7} - 1.5 \cdot 10^{8}$ |
| $D_{TCE,0}$ (cm$^2$·day$^{-1}$) | 0.83 [d] | - |
| $D_{GEO,0}$ (cm$^2$·day$^{-1}$) | 0.24 [e] | 0.06 – 0.72 |
| $C_{TCE,sat}$ (mmol·L$^{-1}$) | 8.4 | - |
| $\theta$ (-) | 0.38 [d] | - |

[a] Parameter value adjusted to fit the experimental results.

[b] Parameter value experimentally determined or optimized from a previous batch experiment (Philips et al., 2013).

[c] Parameter value optimized previously (Haest et al., 2010).

[d] Parameter value experimentally determined previously (Philips et al., 2011).

[e] Parameter value experimentally determined in this study.

[f] For most parameters, the intervals were chosen as a factor 3 around the parameter value, while for $EC_{50}$ and $b_i$ an acceptable interval around the parameter was chosen.

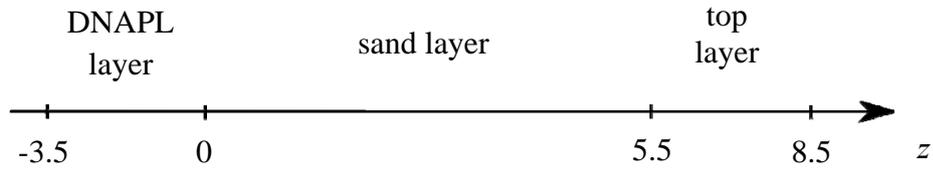
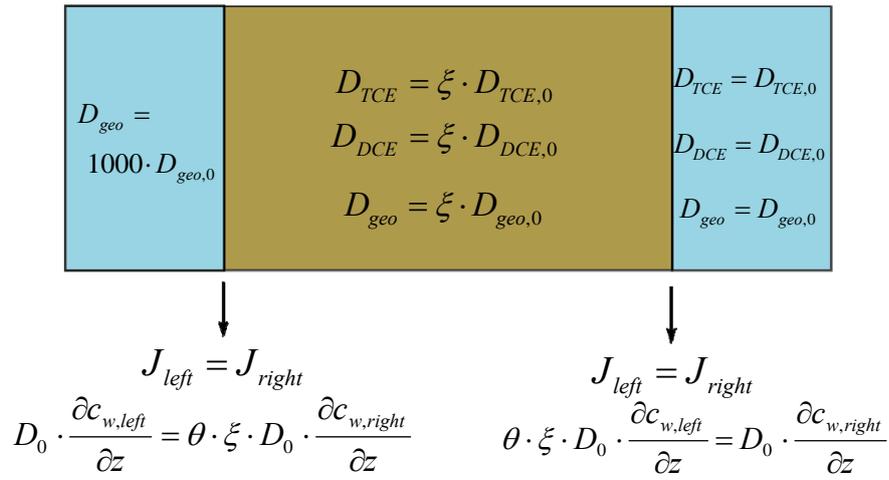
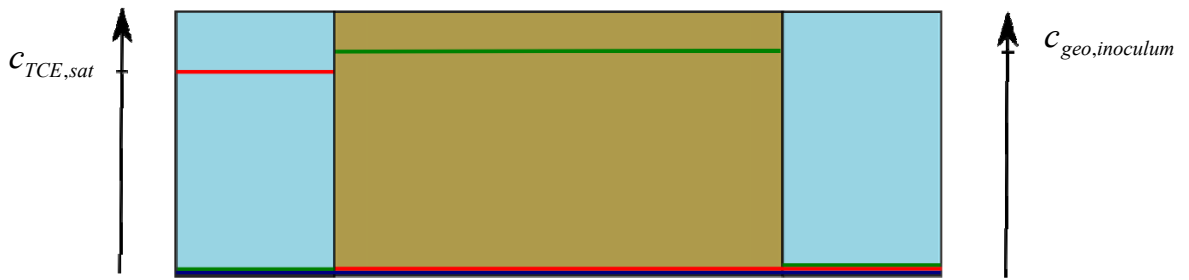
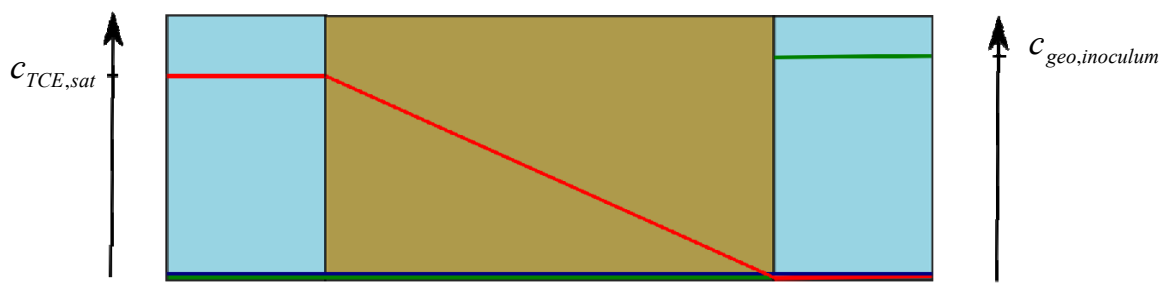
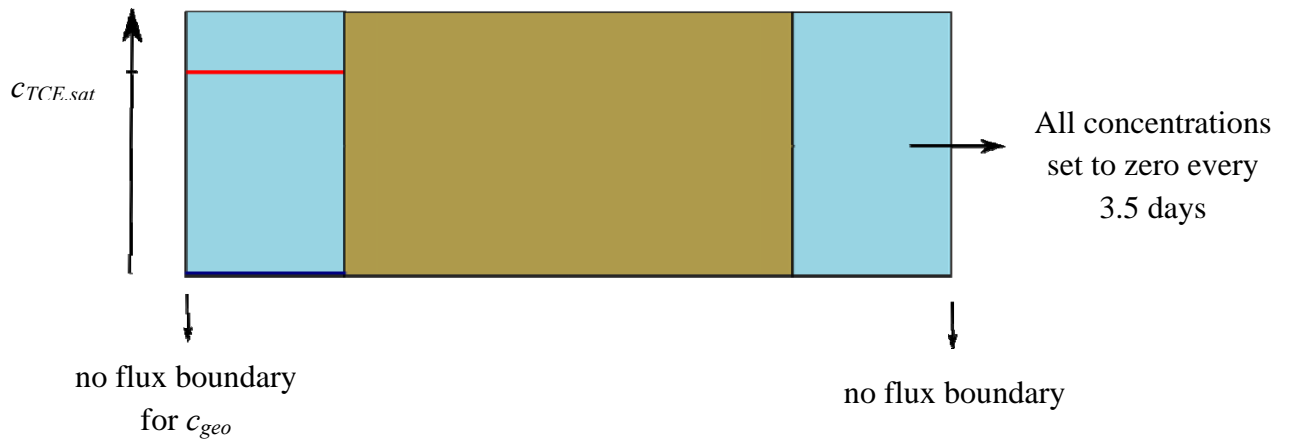

S3

Figure S1: Conditions used to solve the diffusion-motility model.

A: Diffusion and random motility coefficients ($D$) in the central sand layer were adjusted from the coefficients in water ($D_0$) using the tortuosity $\xi$. The mixing in the DNAPL layer was modeled by setting the random motility coefficient 1000 times higher than that in water. Equal diffusion and motility fluxes ($J$) were assumed on the boundaries between different layers, with $c_w$ the concentration or cell density in the water or pore water and $\theta$ the porosity.

B: Initial conditions for sand layer inoculation. The *Geobacter* cell density in the pore water of the central layer was set to one of the different inoculation levels applied ($c_{geo,inoculum}$). Cell densities in the DNAPL and top layer were set zero. The TCE concentration in the DNAPL layer was set to the saturated TCE concentration ($c_{TCE,sat}$). The TCE concentrations throughout the sand and top layer were set zero. All DCE concentrations were set zero.

C: Initial conditions for top layer inoculation. The *Geobacter* cell density in the top layer was set to one of the different inoculation levels applied ($c_{geo,inoculum}$). Zero cell densities were set throughout the sand and DNAPL layer. The TCE concentration in the DNAPL layer was set to the saturated TCE concentration ($c_{TCE,sat}$). The TCE concentration in the sand layer was set to the linearly decreasing steady-state abiotic TCE concentration profile (Philips et al., 2011). A zero TCE concentration was set for the top layer. All DCE concentrations were set zero.

D: Boundary conditions for all diffusion-cells. At all times, the TCE and DCE concentration in the DNAPL layer were set to the saturated TCE concentration ($c_{TCE,sat}$) and zero, respectively. The model incorporated motility of *Geobacter* through the DNAPL layer, the bottom of the diffusion-cell was set as a zero flux boundary for the *Geobacter* cell densities. The model incorporated diffusion of TCE and DCE and motility of *Geobacter* through the top layer, the top of the diffusion-cell was set as a zero flux boundary. Every 3.5 days, all concentrations and cell densities in the top layer were reset to zero to model the top layer refreshment.

Red lines: TCE concentrations, blue lines: DCE concentrations, green lines: *Geobacter* cell densities.

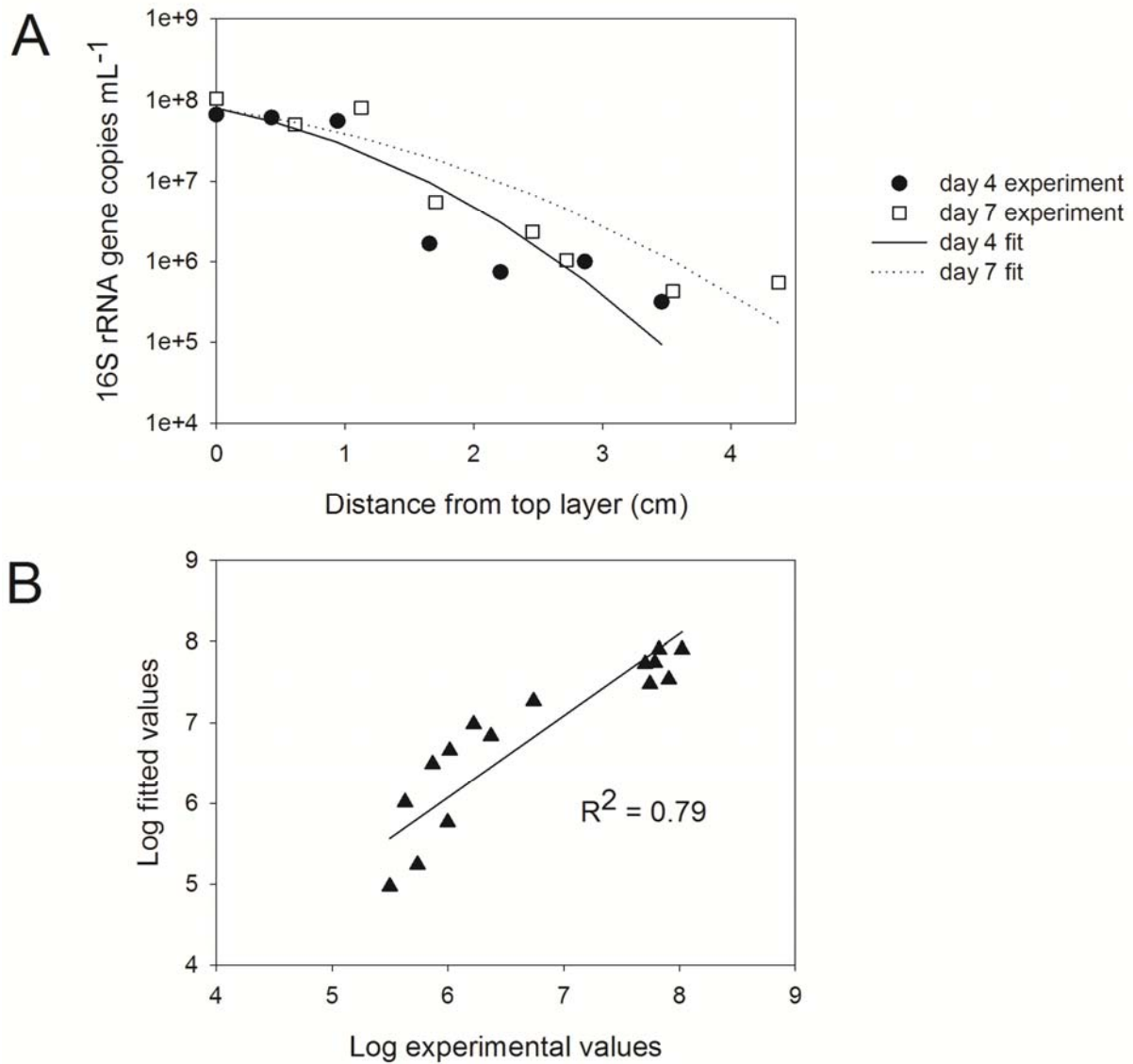

Figure S2: The motility experiment inoculated an aqueous layer on top of a sterile sand layer. (A) *Geobacter* 16S rRNA gene copies were measured after 4 and 7 days and were used to fit Equation 1 in order to determine an effective random motility coefficient for *Geobacter* in the sand. (B) A coefficient of determination ($R^2$) of 0.79 was obtained between log transformed observed and predicted cell densities.

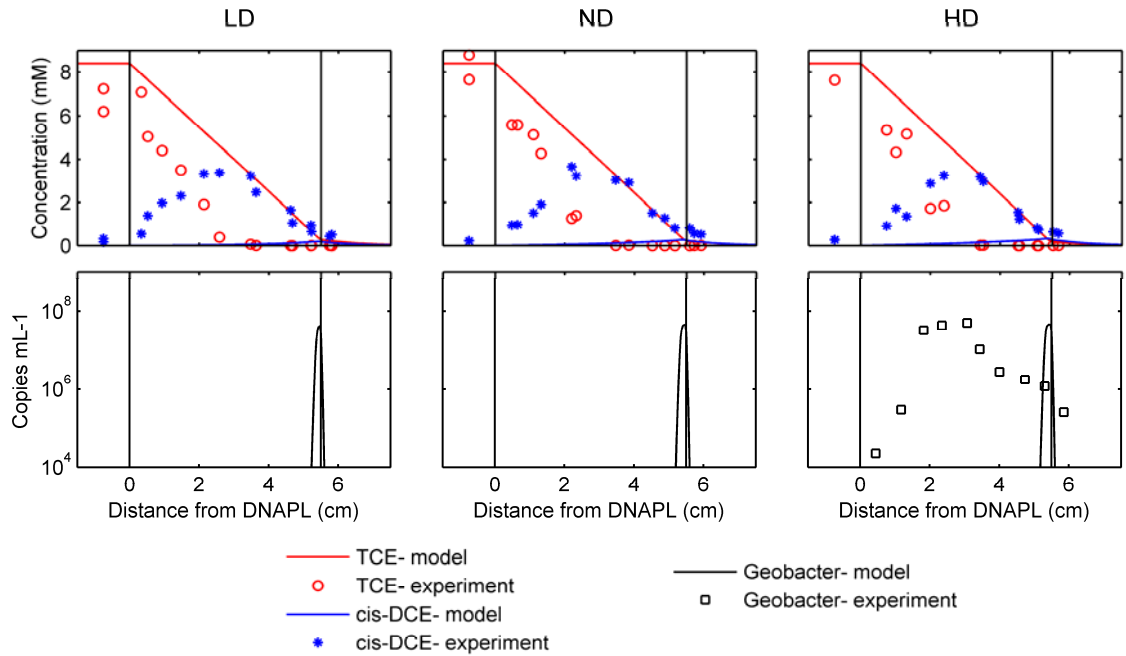

Figure S3: Outcome of the diffusion-motility model assuming immotile *Geobacter* cells versus the experimental TCE and *cis*-DCE concentrations and *Geobacter* 16S rRNA gene copy numbers 19 days after inoculation of the top layer, using three different inoculation densities, i.e. LD: $4·10^5$, ND: $4·10^6$ and HD: $4·10^7$ *Geobacter* 16S rRNA gene copies per mL. Immotile colloids of the size of a bacterium have a Brownian diffusion coefficient which is three orders of magnitude lower than the random motility coefficient for *Geobacter* reported in this study (Ford and Harvey, 2007). As such, the case of immotile *Geobacter* cells was modeled by setting the value for $D_{GEO,0}$ $10^3$ times lower than given in Table S1.

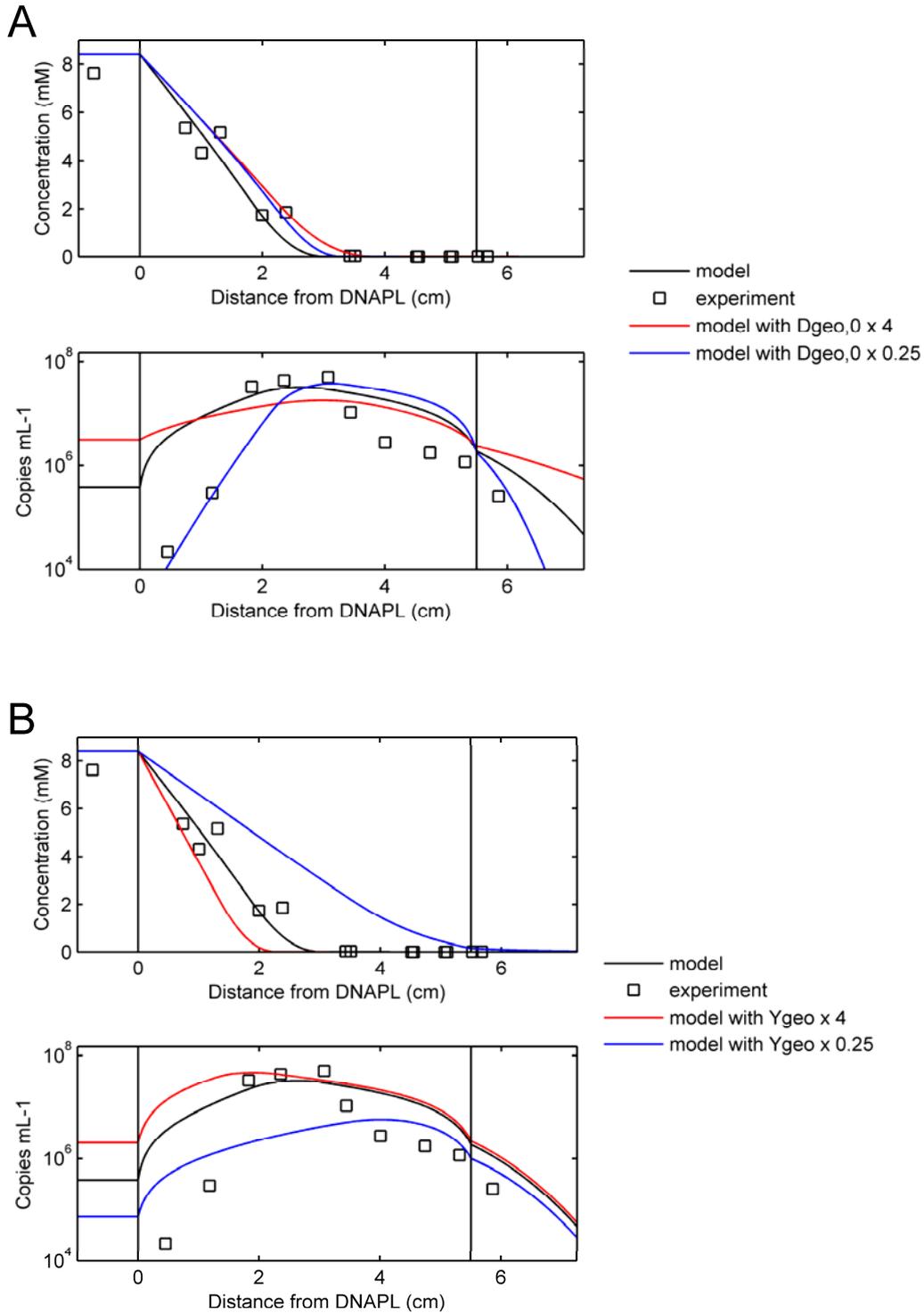

Figure S4: Effect of a 4 times change in parameter value of $D_{GEO,0}$ (A) and $Y_{GEO}$ (B) on the modeled TCE concentrations and *Geobacter* cell densities 19 days after inoculation of the top layer with $4·10^7$ *Geobacter* 16S rRNA gene copies per mL.

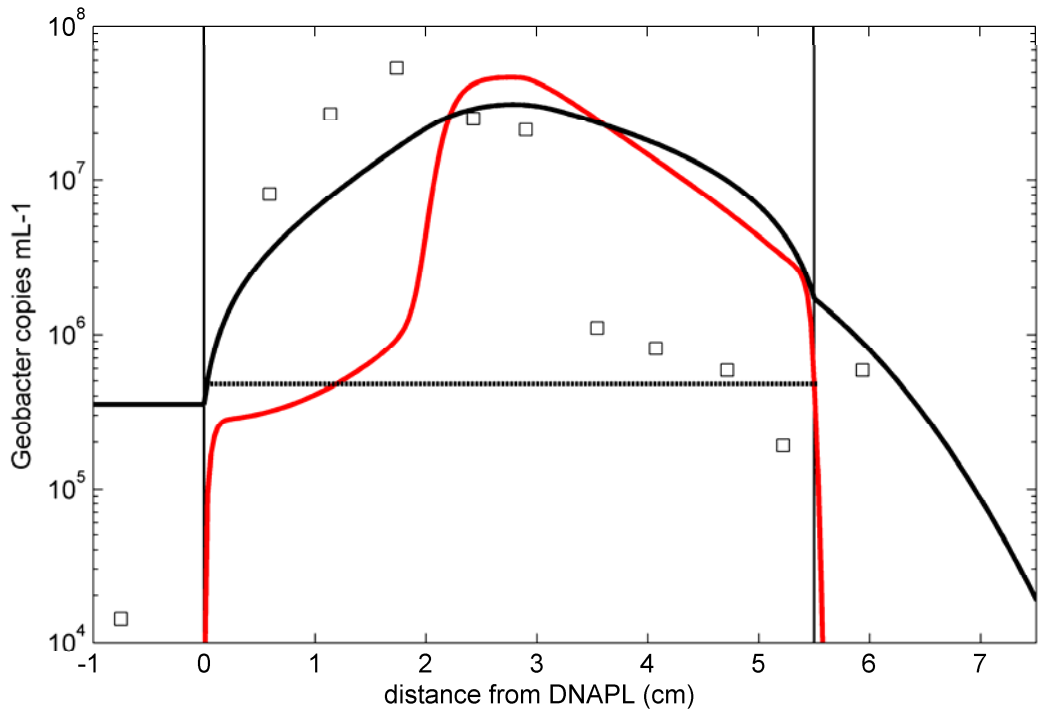

Figure S5: Outcome of the diffusion-motility model using the parameters as in Table S1 (black line) and of the model assuming immotile *Geobacter* cells (red line) versus experimental *Geobacter* 16S rRNA gene copy numbers 19 days after inoculation of the sand layer with an inoculation density of $4 \cdot 10^5$ *Geobacter* 16S rRNA gene copies per mL (indicated by dotted line). The case of immotile *Geobacter* cells was modeled by setting the value for $D_{GEO,0}$ $10^3$ times lower than given in Table S1.